\documentclass[reprint,superscriptaddress,nofootinbib,amsmath,amssymb,physrev,nolongbibliography]{revtex4-2}

\pdfoutput=1
\usepackage{float}
\usepackage{xcolor}
\usepackage{graphicx}
\usepackage{dcolumn}
\usepackage{bm,lipsum}
\usepackage{siunitx}
\usepackage[colorlinks=true, linkcolor=blue, citecolor=blue, urlcolor=blue]{hyperref}
\newcommand{\ZIB}{Zuse Institute Berlin, 14195 Berlin, Germany}
\newcommand{\JCM}{JCMwave GmbH, 14050 Berlin, Germany}

\usepackage{graphicx}
\usepackage{booktabs}
\usepackage{tabularx}
\usepackage{xcolor,colortbl}
\usepackage[T1]{fontenc}

\usepackage[normalem]{ulem}

\begin{document}

\title{Resonance modes in microstructured photonic waveguides: Efficient and accurate computation based on AAA rational approximation}
\author{Felix Binkowski}
\affiliation{\ZIB}
\author{Fridtjof Betz}
\affiliation{\ZIB}
\author{Martin Hammerschmidt}
\affiliation{\JCM}
\author{Lin Zschiedrich}
\affiliation{\JCM}
\author{Sven Burger}
\affiliation{\ZIB}
\affiliation{\JCM}

\begin{abstract}
We present a framework for the efficient and accurate computation of resonance modes in photonic waveguides. The framework is based on AAA rational approximation with the application of special light sources. It allows one to calculate only relevant modes, such as the fundamental resonance modes localized in the central core of the waveguides. We demonstrate the framework using an example from the literature, a hollow-core photonic crystal fiber. This waveguide supports many other modes, such as cladding modes and higher-order modes. These nonrelevant modes are not calculated, so that challenging post-processing with mode filtering is not required.
\end{abstract}

\maketitle
\section{Introduction}
Resonance effects localize optical fields in dielectric fibers and other waveguides and allow for well-defined propagation of light over large distances. Application fields include communication technology~\cite{Kao_1966}, nonlinear optics~\cite{Russell_NatPhot_2014}, sensing~\cite{Ritari_2004}, and imaging~\cite{Wen_2023}. Resonances are the solutions to the source-free Maxwell's equations with open boundary conditions and they are given by electromagnetic fields, the so-called resonance modes, with complex-valued eigenvalues~\cite{Lalanne_QNMReview_2018}. Numerical methods are used to compute the resonances~\cite{Lalanne_QNM_Benchmark_2018,Demesy_ComputPhysComm_2020}, where often many modes of different types are calculated within a dense spectrum.

The functionality of photonic devices based on waveguides is driven by the so-called fundamental resonance modes. These modes are characterized by a localization of the electromagnetic field energy in the central core of the waveguides, which enables low-loss guidance of the light. Furthermore, microstructuring~\cite{Birks_1997,White_OptLett_2002} of the waveguides leads to modes that are localized in the cladding of the systems~\cite{Eggleton_2000}, and fibers with a hollow core support higher-order modes~\cite{Uebel_OL_2016}. Such types of modes are often nonrelevant and make the calculation of the fundamental resonance modes a challenge, as a large number of modes must be calculated and the fundamental resonance modes must then be selected by post-processing. Another challenge is that the eigenvalues of the fundamental resonance modes can have extremely small imaginary parts compared to the real parts due to the low losses~\cite{Roberts_2005,Pearce2007oe}. This leads to very high demands on the numerical accuracy. Therefore, there is a need for approaches that can calculate the fundamental resonance modes of photonic waveguides efficiently and accurately.

Rational approximation is an effective approach for investigating resonant photonic systems. The resulting approximations give the poles and other key figures of the corresponding photonic response functions. The AAA algorithm~\cite{Nakatsukasa_2018_AAA} is a powerful tool for rational approximation. It can be used for the approximation of nonlinear eigenproblems, where the resulting rational eigenproblems can then be solved with suitable numerical methods~\cite{Elsworth_2019,Lietaert_2021,Guettel_weightedAAA_2022}. Recently, approaches have been presented, where rational approximation with~\cite{Bruno_AAA_2024,Betz_LPOR_2024} and without~\cite{Pradovera_PAMM_2023} applying the AAA algorithm is used to directly solve eigenproblems, i.e., to compute the resonance modes associated to the poles of the response functions of interest.

\begin{figure*}
\includegraphics[width=0.98\textwidth]{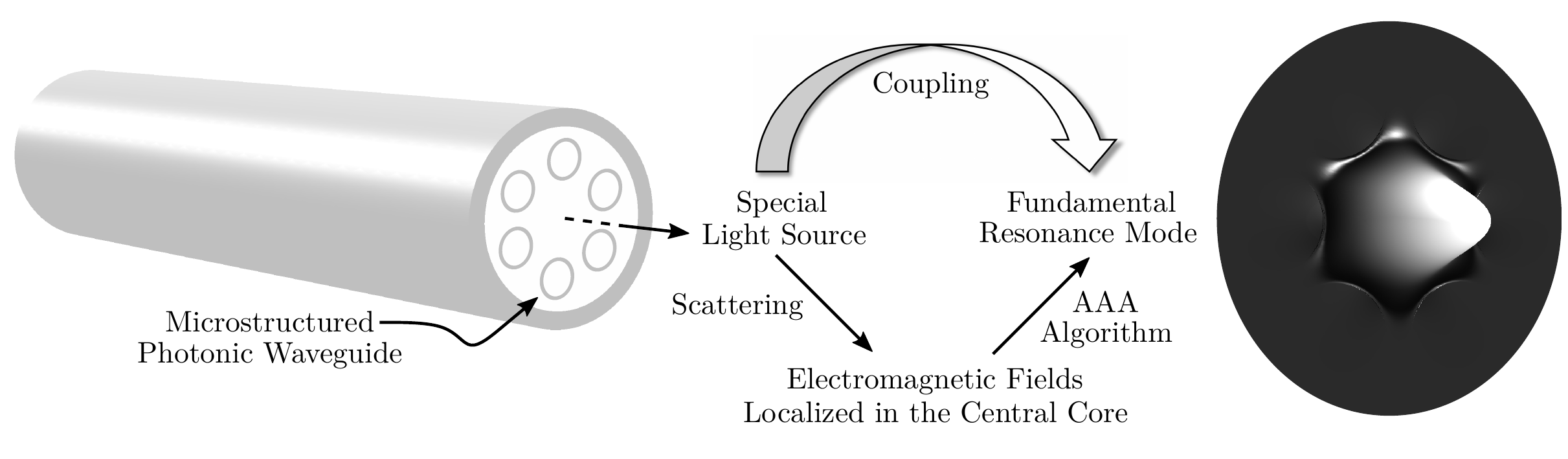}
\caption{\label{fig1}
Computation of the fundamental resonance mode of a microstructured waveguide. The waveguide is illuminated with a light source located at the center of the system. The resulting scattered electromagnetic fields are superposed based on AAA rational approximation. The light source is chosen so that it couples mainly with the fundamental resonance mode and not with other modes. This special choice of the source enables an efficient and accurate computation of the fundamental resonance mode of the system.
}
\end{figure*}

In this work, we present a framework based on AAA rational approximation to compute the fundamental resonance modes of photonic waveguides. The modes are calculated with the approach proposed in Ref.~\cite{Betz_LPOR_2024} using special light sources. We apply the framework to a hollow-core photonic crystal fiber (HC-PCF), where the microstructuring of the fiber leads to the existence of many cladding modes and the hollow core enables the presence of higher-order modes. The framework allows for an efficient and accurate computation of the fundamental resonance mode, where the other types of modes are not calculated. Challenging post-processing is therefore not necessary. The results are compared with the results obtained by the Arnoldi algorithm, which is a standard tool in the field of computational photonics.

Figure~\ref{fig1} outlines an application example of the framework. The fundamental resonance mode of a photonic waveguide is characterized by a localization of the corresponding electromagnetic field energy in the central core of the waveguide. A specially selected light source, which is located at the center of the system, has a significant coupling with the mode. Application of the AAA algorithm to the fields caused by the source yields the fundamental resonance mode.

\section{Computing resonance modes with the AAA algorithm}
The AAA algorithm~\cite{Nakatsukasa_2018_AAA} gives an approximation of a scalar-valued function $f(z)$ by a rational function $r(z)$ in a barycentric representation. A number $M$ of freely selectable sampling points $z_k \in Z\subseteq\mathbb{C}$ and corresponding function values $f_k = f(z_k)$ are the input for the algorithm. The algorithm greedily adds sampling points $\hat{z}_j$ to a subset $\hat{Z}\subset Z$, together with the corresponding function values $\hat{f}_j$. Then, each iteration within the algorithm leads to a rational approximation $r(z)$ of order $m-1$,
\begin{equation} \label{eq:r}
    r(z) = \frac{n(z)}{d(z)} = \left. \sum_{j=1}^{m}\frac{\hat{w}_j \hat f_j}{z-\hat z_j}\middle/\sum_{j=1}^{m} \frac{\hat{w}_j}{z-\hat z_j} \right.,
\end{equation}
where the weights $\hat{w}_j$ minimize the error 
\begin{equation} \label{eq:lsq}
    \sum_{z_k\in Z\backslash\hat{Z}}|f_k \, d(z_k)-n(z_k)|^2.
\end{equation} 
The least square problem in Eq.~\eqref{eq:lsq} is solved using a singular value decomposition with the constraint $\sum_{j=1}^m |\hat{w}_j|^2 = 1$. The solution of the least square problem requires $m\leq M/2$. The AAA algorithm also directly provides the underlying key figures of the rational approximation, such as the poles $z_n^\mathrm{pole} \in \mathbb{C}$ and the residues $a_n \in \mathbb{C}$.

In the following, we consider a physical vector-valued quantity $\mathbf{f}(z) \in \mathbb{C}^{N}$ which is the solution of the linear system of equations
\begin{equation} \label{eq:LSE}
    \mathbf{A}(z) \mathbf{f}(z) = \mathbf{s}(z),
\end{equation} 
where $\mathbf{A}(z) \in \mathbb{C}^{N\times N}$ is the system matrix and $\mathbf{s}(z) \in \mathbb{C}^{N}$ is an impressed source term. It is further given that $f_k = \mathcal{L}(\mathbf{f}_k)$, where $\mathcal{L} : \mathbb{C}^N \rightarrow \mathbb{C}$ is a linear mapping and $f_k$ are the function values of the scalar-valued function $f(z)$ as introduced above. Then, the AAA algorithm also yields the vector-valued approximation
\begin{equation*} \label{eq:rv}
    \mathbf{r}(z) = \left. \sum_{j=1}^{m}\frac{\hat{w}_j \hat{\mathbf{f}}_j}{z-\hat z_j}\middle/ d(z) \right. \approx \mathbf{f}(z)
\end{equation*}
and the corresponding vector-valued residue
\begin{equation} \label{eq:eigenvector}
    \mathbf{a}_n = \sum_{j=1}^m \left[\frac{\hat{w}_j}{z_n^\mathrm{pole}-\hat{z}_j} \middle/ \frac{\partial d}{\partial z}\left(z_n^\mathrm{pole}\right)\right]\hat{\mathbf{f}}_j,
\end{equation}
where the vectors $\hat{\mathbf{f}}_j$ are defined by the relation $\hat{f}_j = \mathcal{L}(\hat{\mathbf{f}}_j)$. Note that the weights $\hat{w}_j$ and poles $z_n^\mathrm{pole}$ in Eq.~\eqref{eq:eigenvector} are the same as those used for the scalar-valued rational approximation $r(z)$ from Eq.~\eqref{eq:r}.

When a pole $z_n^\mathrm{pole}$ has a significant influence on the rational approximation $r(z)$, then we assume that $\mathbf{a}_n$ and $z_n^\mathrm{pole}$ are a good approximation to an eigenpair of the corresponding nonlinear eigenproblem, i.e., $\mathbf{A}(z_n^\mathrm{pole})\mathbf{a}_n \approx 0$. This means that the resonance mode $\mathbf{a}_n$ corresponding to the eigenvalue $z_n^\mathrm{pole}$ has a significant coupling with the source term $\mathbf{s}(z)$~\cite{Betz_LPOR_2024}.

Note that, to compute the solution $\mathbf{f}(z)$ of Eq.~\eqref{eq:LSE}, any black-box solver can be used. For the proposed approach, only access to the solution is required, i.e., the way in which the solution is calculated is not taken into account.

\section{Application}
\begin{figure}
\includegraphics[width=0.49\textwidth]{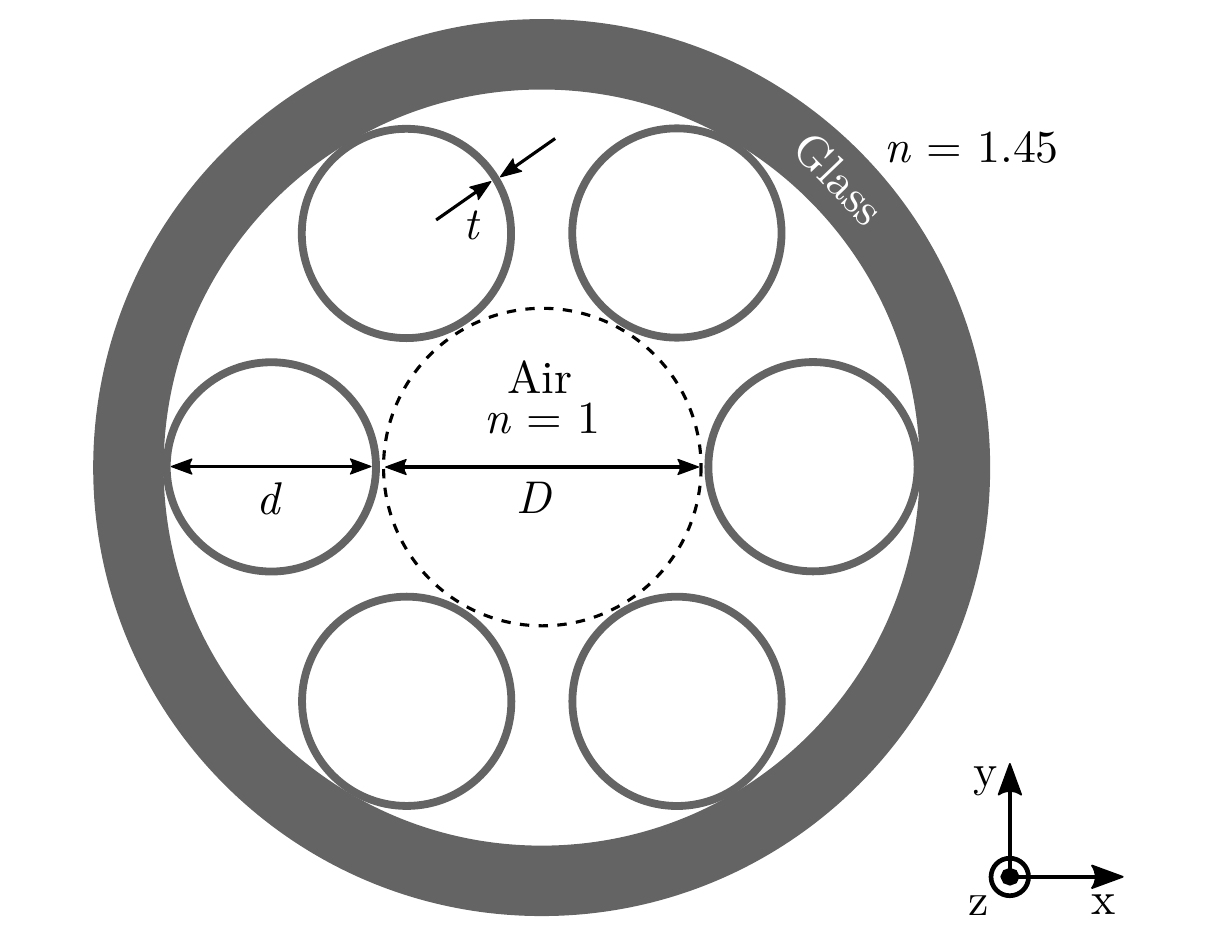}
\caption{\label{fig2}
Sketch of the HC-PCF presented in Ref.~\cite{Uebel_OL_2016}. The diameter of the central hollow core is $D=30\,\mu\mathrm{m}$. The core is encircled by six nontouching glass rings with a wall thickness of $t = D \times 0.01$. The inner diameter of the glass rings is $d = D \times 0.68$. The HC-PCF is coated with thick-walled glass. The vacuum wavelength is set to $\lambda_0 = 1500 \mathrm{nm}$.
}
\end{figure}

We apply the approach presented to compute the fundamental resonance mode of a photonic waveguide from the literature, the HC-PCF introduced in Ref.~\cite{Uebel_OL_2016}. The system is sketched in~Fig.~\ref{fig2}. The longitudinal axis of the system is along the $z$-direction, and this dimension is much larger than the diameter of the cross-section of the system. Based on this, we model the HC-PCF with an infinite length in $z$-direction and we assume a harmonic dependence of the scattered electric fields $\mathbf{E}(x,y,z) \in \mathbb{C}^3$ and impressed current densities $\mathbf{J}(x,y,z) \in \mathbb{C}^3$ on the $z$-coordinate, i.e., $\mathbf{E}(x,y,z) = \mathbf{E}(x,y)e^{i k_z z}$ and $\mathbf{J}(x,y,z) = \mathbf{J}(x,y)e^{i k_z z}$, where $k_z \in \mathbb{C}$ is the propagation constant. With this, in the steady-state regime, light scattering in the system can be described by the time-harmonic Maxwell's equation in second-order form,
\begin{align}
\begin{split}
	\nabla_{k_z} \hspace{-0.1cm} \times \hspace{-0.05cm} \mu^{-1} 
	&\nabla_{k_z} \hspace{-0.1cm} \times \hspace{-0.05cm} \mathbf{E}(x,y) -
	\omega_0^2\epsilon \mathbf{E}(x,y)  = 
	i\omega_0\mathbf{J}(x,y), \label{eq:Maxwell}
 \end{split}
\end{align}
where $\nabla_{k_z} = (\partial_x,\partial_y,ik_z)^T$. 
The material is characterized by the complex-valued permittivity and permeability tensors $\epsilon(x,y)$ and $\mu(x,y)$,  respectively.
The angular frequency $\omega_0 = 2 \pi c / \lambda_0$ is a fixed parameter, where $\lambda_0$ is the vacuum wavelength.

The electric field in the HC-PCF is excited with a singular electric current density on a line along the $z$-direction,
\begin{align}
    \mathbf{J}(x,y) = \mathbf{j}\delta((x,y)-(x_0,y_0))e^{-i k_z z_0}, \nonumber
\end{align}
where $(x_0,y_0,z_0)$ is the position of the line source, $\delta$ is the Dirac delta distribution, and $\mathbf{j}$ is a constant strength vector. Since the fundamental resonance mode is localized in the central hollow core of the HC-PCF, we place the line source at the center of the system. We further consider an $x$-polarized line source, i.e., $\mathbf{j} = |\mathbf{j}| \times (1,0,0)^T$.

In order to solve Eq.~\eqref{eq:Maxwell}, we use the solver JCMsuite, which is based on the finite element method (FEM).
The thick-walled glass cladding of the HC-PCF is modeled to extend to infinity, i.e., open boundaries realized by perfectly matched layers are applied. We further exploit the double mirror symmetry of the system in the numerical implementations. Numerical convergence with respect to the FEM parameters is ensured. The corresponding settings can be found in the data publication~\cite{Binkowski_SourceCode_AAA_HCPCF}.

\subsection{Reference solutions from the Arnoldi algorithm}
Reference solutions are obtained by applying the Arnoldi algorithm~\cite{Saad_Book_NumMeth_Eig_2011,Lalanne_QNM_Benchmark_2018,Demesy_ComputPhysComm_2020} within JCMsuite to the eigenproblem
\begin{align}
\begin{split}
	\nabla_{k_{z,n}} \hspace{-0.1cm} \times \hspace{-0.05cm} \mu^{-1} 
	&\nabla_{k_{z,n}} \hspace{-0.1cm} \times \hspace{-0.05cm} \mathbf{E}_n(x,y) -
	\omega_0^2\epsilon \mathbf{E}_n(x,y)  = 
	0, \label{eq:Maxwell_sourcefree}
 \end{split}
\end{align}
i.e., the source-free form of Eq.~\eqref{eq:Maxwell}. The Arnoldi algorithm requires a guess value for the eigenvalues $k_{z,n}$, where $\omega_0$ is fixed. Then, it iteratively calculates a selected number of eigenvalues closest to the guess value, together with the corresponding resonance modes $\mathbf{E}_n$. The boundary conditions on the symmetry axes are chosen such that the polarization of the modes matches the polarization of the line source used for the AAA algorithm. In the following, for a traditional notation, the eigenvalues $k_{z,n}$ are given in the form of effective refractive indices $n_n^\mathrm{eff} = k_{z,n}/k_0$, where $k_0 = 2 \pi / \lambda_0$.

\subsection{Rational approximation and eigenvalues}
The optical response resulting from the illumination of the HC-PCF is investigated by the quantity $y^T\mathbf{E}_x \in \mathbb{C}$, where $y \in \mathbb{R}^{m}$ is a random vector~\cite{Elsworth_2019} with a uniform distribution in the interval $(-1, 1)$ and $\mathbf{E}_x \in \mathbb{C}^{m}$ are the $x$-components of the electric field $\mathbf{E}$, which is determined on an equidistantly spaced cartesian grid in one of the four mirror symmetry planes. The circular computational domain with 100 points in $x$ and $y$-direction leads to $m=7787$ evaluation points. The electric field $\mathbf{E}$ is obtained by solving the scattering problem given by Eq.~\eqref{eq:Maxwell} at chosen sampling points $n_j^\mathrm{eff}$.

\begin{figure}[]
\includegraphics[width=0.49\textwidth]{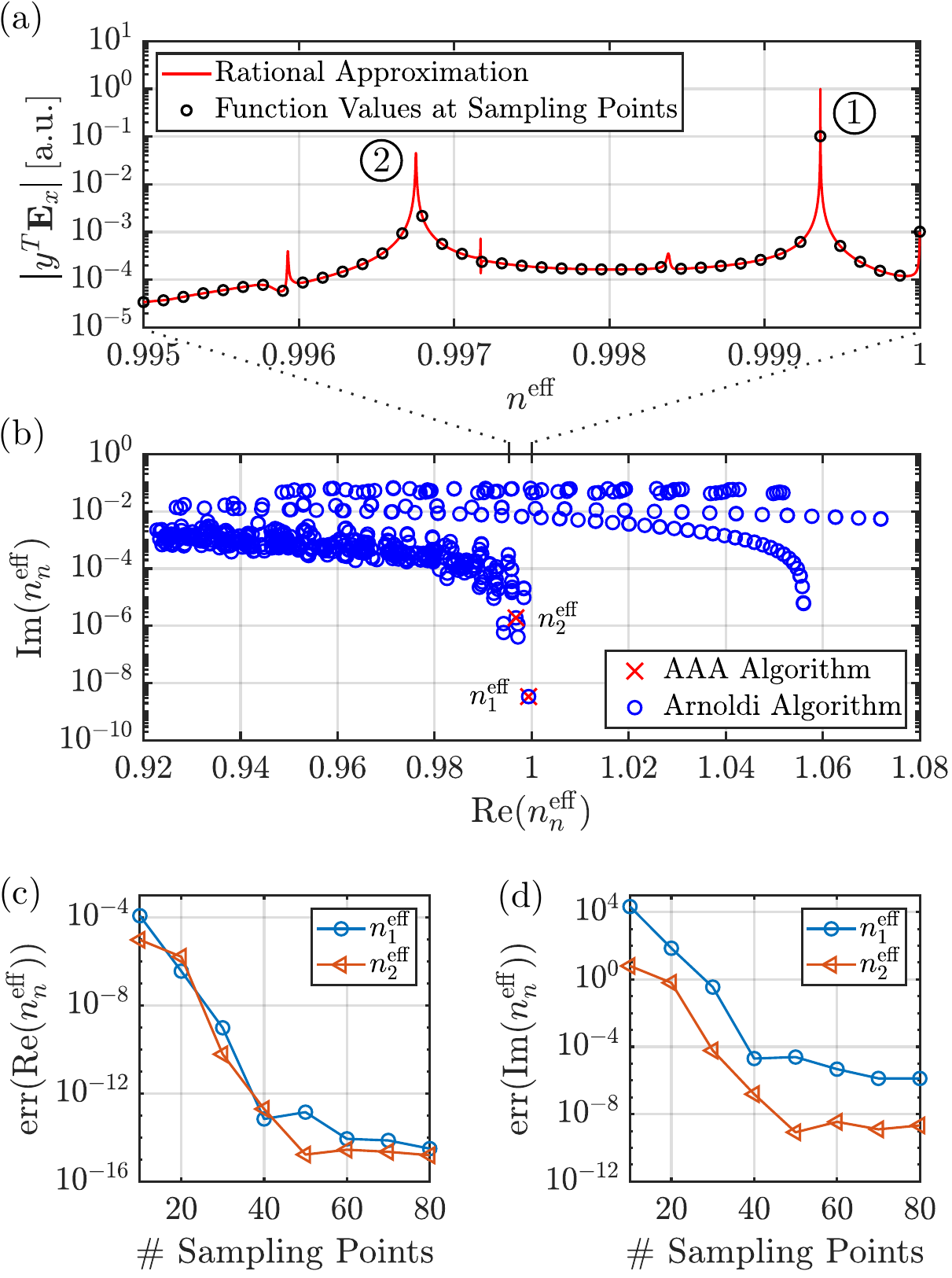}
\caption{\label{fig3}
Illumination of the HC-PCF sketched in Fig.~\ref{fig2} with an $x$-polarized line source located at the center of the system. Application of the AAA algorithm to the quantity $y^T\mathbf{E}_x \in \mathbb{C}$, where $y \in \mathbb{R}^{m}$ is a random vector and $\mathbf{E}_x \in \mathbb{C}^{m}$ are the $x$-components of the electric field $\mathbf{E}$, which is evaluated at $m$ spatial points in the computational domain.
(a)~Absolute values of the rational approximation based on $40$ equidistantly spaced sampling points $n_j^\mathrm{eff} \in [0.995, 1]$. The absolute values of the function values at the $40$ sampling points are also shown.
(b)~Eigenvalues $n_1^\mathrm{eff}$ and $n_2^\mathrm{eff}$ resulting from the rational approximation and reference eigenvalues computed by the Arnoldi algorithm. 
(c)~Relative errors $\mathrm{err}(\mathrm{Re}(n_n^\mathrm{eff})) = |(\mathrm{Re}(n_n^\mathrm{eff})-\mathrm{Re}(n_{n,\mathrm{ref}}^\mathrm{eff}))/\mathrm{Re}(n_{n,\mathrm{ref}}^\mathrm{eff})|$ over number of equidistantly spaced sampling points in the interval $n^\mathrm{eff} \in [0.995, 1]$, where the reference solutions $n_{n,\mathrm{ref}}^\mathrm{eff}$ are computed by the Arnoldi algorithm.
(d)~Relative errors of the imaginary parts of the eigenvalues.
}
\end{figure}

\begin{figure*}
\includegraphics[width=0.98\textwidth]{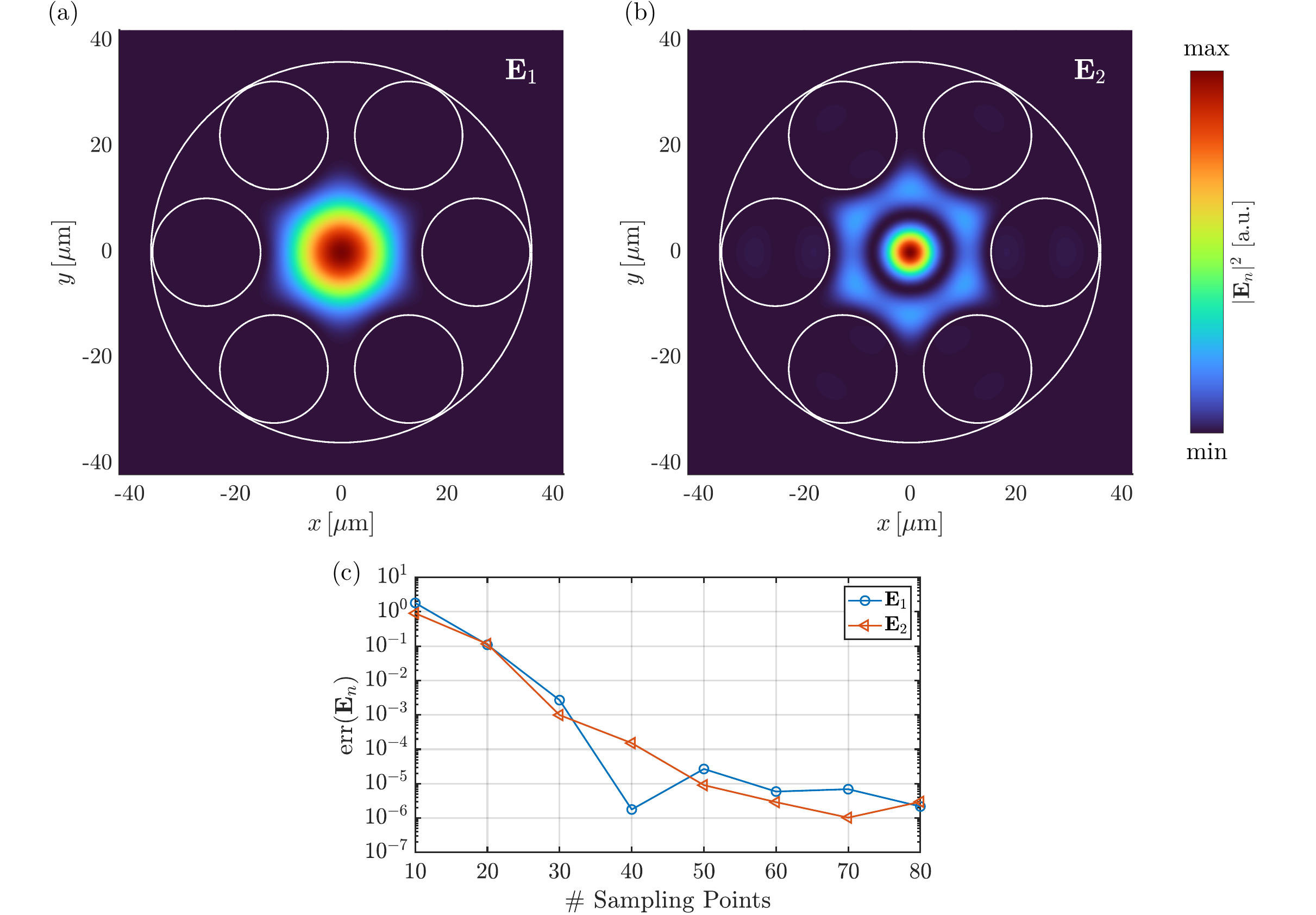}
\caption{\label{fig4}
Computation of resonance modes of the HC-PCF sketched in Fig.~\ref{fig2} using the AAA algorithm.
(a)~Electric field intensity of the fundamental resonance mode $\mathbf{E}_1$ corresponding to the eigenvalue $n^\mathrm{eff}_1$.
(b)~Electric field intensity of the resonance mode $\mathbf{E}_2$ corresponding to the eigenvalue $n^\mathrm{eff}_2$. 
The modes are scaled differently.
(c)~Relative error $\mathrm{err}(\mathbf{E}_n) = \left\Vert \mathbf{E}_n-\mathbf{E}_{n,\mathrm{ref}}\right \Vert/\left\Vert\mathbf{E}_{n,\mathrm{ref}}\right \Vert$ over number of equidistantly spaced sampling points in the interval $n^\mathrm{eff} \in [0.995, 1]$ for the AAA algorithm, where the reference solutions $\mathbf{E}_{n,\mathrm{ref}}$ are computed by the Arnoldi algorithm. The norm $\Vert \cdot \Vert$ is defined as the square root of the electric field energy in the computational domain. The resonance modes $\mathbf{E}_n$ and $\mathbf{E}_{n,\mathrm{ref}}$ are normalized such that their $x$-components at the center of the system are equal.
}
\end{figure*}

We apply the AAA algorithm to $y^T\mathbf{E}_x$ computed at $40$ equidistantly spaced sampling points $n_j^\mathrm{eff} \in [0.995, 1]$. The absolute values of the resulting rational approximation and of the underlying function values are shown in Fig.~\ref{fig3}(a). In Fig.~\ref{fig3}(b), the eigenvalues of the system are presented. We  show $512$ eigenvalues computed by the Arnoldi algorithm as reference solutions, where a guess eigenvalue of $0.9975$ is chosen. The eigenvalues correspond to cladding modes, higher-order modes, and also to the fundamental resonance mode. In contrast, the application of the AAA algorithm with the special line source only yields relevant eigenvalues. They belong to the rational approximation shown in Fig.~\ref{fig3}(a) and they are selected based on the two significant peaks, marked with (1) and (2). These eigenvalues are given by $n_1^\mathrm{eff} = 0.9993596784939 + 0.000000003376i$ and $n_2^\mathrm{eff} =  0.996754264645 + 0.00000190093i$. The eigenvalue $n_1^\mathrm{eff}$ has the smallest imaginary part of all eigenvalues in the chosen range for the effective refractive index. The other eigenvalues~\cite{Betz_LPOR_2024} corresponding to the rational approximation, which arise due to the other peaks, the background continuum, or the eigenvalues outside the chosen range of the effective refractive index, and further details on the settings for the algorithm can be found in the data publication~\cite{Binkowski_SourceCode_AAA_HCPCF}.

Figure~\ref{fig3}(c) shows the relative errors of the real parts of the eigenvalues $n_1^\mathrm{eff}$ and $n_2^\mathrm{eff}$ over the number of sampling points for the AAA algorithm. For both eigenvalues, convergence up to errors smaller than $10^{-14}$ can be observed. Figure~\ref{fig3}(d) shows the relative errors of the imaginary parts, where errors smaller than $10^{-5}$ and smaller than $10^{-8}$ for $n_1^\mathrm{eff}$ and $n_2^\mathrm{eff}$, respectively, are achieved. The limitation of the accuracy can be attributed to the accuracy of the FEM scattering solver.

Note that, since the framework essentially relies on solving scattering problems, the incorporation of sensitivities based on algorithmic differentiation is possible~\cite{Binkowski_CommunPhys_2022,Betz_LPOR_2024}. This means that also the sensitivities of the eigenvalues with respect to the system parameters are available with negligible computational costs.

\subsection{Fundamental resonance mode}
To compute the fundamental resonance mode of the HC-PCF, we apply Eq.~\eqref{eq:eigenvector} using the weights, sampling points, and eigenvalues corresponding to the rational approximation from Fig.~\ref{fig3}(a). The matrix $\mathbf{A}(z)$ from Eq.~\eqref{eq:LSE} is the FEM system matrix, the vector $\mathbf{s}(z)$ corresponds to the impressed line source, and the vector $\mathbf{f}(z)$ is the scattered electric field in a finite-dimensional FEM basis. This means that, for the vectors $\hat{\mathbf{f}}_j$ in Eq.~\eqref{eq:eigenvector}, the FEM coefficient vectors corresponding to the electric fields $\mathbf{E}$ are used.

Figure~\ref{fig4}(a) and (b) show the electric field intensities of the resonance modes $\mathbf{E}_1$ and $\mathbf{E}_2$ corresponding to $n_1^\mathrm{eff}$ and $n_1^\mathrm{eff}$, respectively. We identify $\mathbf{E}_1$ as the fundamental resonance mode, as it has no nodal lines within the field pattern. Both modes are localized in the central hollow core of the HC-PCF. The chosen line source for the illumination of the system is located at the center and therefore exhibits a significant coupling with the modes. Figure~\ref{fig4}(c) shows the relative errors of the two modes over the number of sampling points for the AAA algorithm. Convergence up to errors smaller than $10^{-5}$ can be observed.

Note that, in this work, we choose real-valued sampling points. The accuracy of the resonance modes and their eigenvalues could be further improved if sampling points in the complex plane are chosen. Further information on this topic can be found, e.g., in Ref.~\cite{Betz_LPOR_2024}, where an approach based on adaptive sampling near the physical meaningful eigenvalues is presented.

We further point out that other types of sources could also be used for the framework presented, e.g., a combination of multiple line sources, the field distribution of the fundamental resonance mode of a single-mode fiber, or a Gaussian beam.

\section{Conclusion}
We presented a framework based on AAA rational approximation to compute the fundamental resonance modes of photonic waveguides. The framework was applied to an example from the literature, a HC-PCF supporting many cladding and higher-order modes. The scattering solutions at the sampling points for the AAA algorithm were superposed, reusing the poles and weights belonging to a corresponding scalar-valued rational approximation. The coupling of the underlying light source with the fundamental resonance mode of the HC-PCF enables an efficient and accurate computation of the mode.

The results were compared with the eigenpairs obtained using the Arnoldi algorithm. The Arnoldi algorithm solves the eigenproblem directly, i.e., without a source term, and it calculates all resonance modes together with the associated eigenvalues closest to a guess eigenvalue. The AAA algorithm is based on solving scattering problems at chosen sampling points, i.e., with using a source term, and, therefore, only calculates the modes that have a significant coupling with the applied source.

The framework presented is beneficial when the system of interest supports modes that are much more relevant than other modes, such as in the case studied in this work. The fundamental resonance mode of the HC-PCF investigated has an intensity maximum at the center of the system and it can be efficiently excited by a singular line source located at the center. Many of the additional resonance modes of the HC-PCF are modes that are localized in the cladding of the fiber or higher-order modes. They have an insignificant coupling with the line source, i.e., they are not calculated when using the AAA algorithm. This means that challenging post-processing with mode filtering, as may be necessary with the Arnoldi algorithm, can be avoided.

\section*{Data availability} 
Supplementary data tables and source code for the numerical experiments for this work can be found in the open access data publication~\cite{Binkowski_SourceCode_AAA_HCPCF}.

\section*{Acknowledgments}
We gratefully acknowledge discussions with Lloyd N.~Trefethen. 
We acknowledge funding by the Deutsche Forschungsgemeinschaft (DFG, German Research Foundation) under Germany's Excellence Strategy - The Berlin Mathematics Research Center MATH+ (EXC-2046/1, project ID: 390685689) and by the German Federal Ministry of Education and Research (BMBF Forschungscampus MODAL, project 05M20ZBM).


\begin{thebibliography}{23}%
	\makeatletter
	\providecommand \@ifxundefined [1]{%
		\@ifx{#1\undefined}
	}%
	\providecommand \@ifnum [1]{%
		\ifnum #1\expandafter \@firstoftwo
		\else \expandafter \@secondoftwo
		\fi
	}%
	\providecommand \@ifx [1]{%
		\ifx #1\expandafter \@firstoftwo
		\else \expandafter \@secondoftwo
		\fi
	}%
	\providecommand \natexlab [1]{#1}%
	\providecommand \enquote  [1]{``#1''}%
	\providecommand \bibnamefont  [1]{#1}%
	\providecommand \bibfnamefont [1]{#1}%
	\providecommand \citenamefont [1]{#1}%
	\providecommand \href@noop [0]{\@secondoftwo}%
	\providecommand \href [0]{\begingroup \@sanitize@url \@href}%
	\providecommand \@href[1]{\@@startlink{#1}\@@href}%
	\providecommand \@@href[1]{\endgroup#1\@@endlink}%
	\providecommand \@sanitize@url [0]{\catcode `\\12\catcode `\$12\catcode
		`\&12\catcode `\#12\catcode `\^12\catcode `\_12\catcode `\%12\relax}%
	\providecommand \@@startlink[1]{}%
	\providecommand \@@endlink[0]{}%
	\providecommand \url  [0]{\begingroup\@sanitize@url \@url }%
	\providecommand \@url [1]{\endgroup\@href {#1}{\urlprefix }}%
	\providecommand \urlprefix  [0]{URL }%
	\providecommand \Eprint [0]{\href }%
	\providecommand \doibase [0]{https://doi.org/}%
	\providecommand \selectlanguage [0]{\@gobble}%
	\providecommand \bibinfo  [0]{\@secondoftwo}%
	\providecommand \bibfield  [0]{\@secondoftwo}%
	\providecommand \translation [1]{[#1]}%
	\providecommand \BibitemOpen [0]{}%
	\providecommand \bibitemStop [0]{}%
	\providecommand \bibitemNoStop [0]{.\EOS\space}%
	\providecommand \EOS [0]{\spacefactor3000\relax}%
	\providecommand \BibitemShut  [1]{\csname bibitem#1\endcsname}%
	\let\auto@bib@innerbib\@empty
	\bibitem [{\citenamefont {Kao}\ and\ \citenamefont {Hockham}(1966)}]{Kao_1966}%
	\BibitemOpen
	\bibfield  {author} {\bibinfo {author} {\bibfnamefont {K.}~\bibnamefont
			{Kao}}\ and\ \bibinfo {author} {\bibfnamefont {G.}~\bibnamefont {Hockham}},\
	}\href {https://doi.org/10.1049/piee.1966.0189} {\bibfield  {journal}
		{\bibinfo  {journal} {Proc. Inst. Electr. Eng.}\ }\textbf {\bibinfo {volume}
			{113}},\ \bibinfo {pages} {1151} (\bibinfo {year} {1966})}\BibitemShut
	{NoStop}%
	\bibitem [{\citenamefont {{P. St. J. Russell}}\ \emph
		{et~al.}(2014)\citenamefont {{P. St. J. Russell}}, \citenamefont {Hölzer},
		\citenamefont {Chang}, \citenamefont {Abdolvand},\ and\ \citenamefont
		{Travers}}]{Russell_NatPhot_2014}%
	\BibitemOpen
	\bibfield  {author} {\bibinfo {author} {\bibnamefont {{P. St. J. Russell}}},
		\bibinfo {author} {\bibfnamefont {P.}~\bibnamefont {Hölzer}}, \bibinfo
		{author} {\bibfnamefont {W.}~\bibnamefont {Chang}}, \bibinfo {author}
		{\bibfnamefont {A.}~\bibnamefont {Abdolvand}},\ and\ \bibinfo {author}
		{\bibfnamefont {J.~C.}\ \bibnamefont {Travers}},\ }\href
	{https://doi.org/10.1038/nphoton.2013.312} {\bibfield  {journal} {\bibinfo
			{journal} {Nat. Photonics}\ }\textbf {\bibinfo {volume} {8}},\ \bibinfo
		{pages} {278} (\bibinfo {year} {2014})}\BibitemShut {NoStop}%
	\bibitem [{\citenamefont {Ritari}\ \emph {et~al.}(2004)\citenamefont {Ritari},
		\citenamefont {Tuominen}, \citenamefont {Ludvigsen}, \citenamefont
		{Petersen}, \citenamefont {S{\o}rensen}, \citenamefont {Hansen},\ and\
		\citenamefont {Simonsen}}]{Ritari_2004}%
	\BibitemOpen
	\bibfield  {author} {\bibinfo {author} {\bibfnamefont {T.}~\bibnamefont
			{Ritari}}, \bibinfo {author} {\bibfnamefont {J.}~\bibnamefont {Tuominen}},
		\bibinfo {author} {\bibfnamefont {H.}~\bibnamefont {Ludvigsen}}, \bibinfo
		{author} {\bibfnamefont {J.~C.}\ \bibnamefont {Petersen}}, \bibinfo {author}
		{\bibfnamefont {T.}~\bibnamefont {S{\o}rensen}}, \bibinfo {author}
		{\bibfnamefont {T.~P.}\ \bibnamefont {Hansen}},\ and\ \bibinfo {author}
		{\bibfnamefont {H.~R.}\ \bibnamefont {Simonsen}},\ }\href
	{https://doi.org/10.1364/OPEX.12.004080} {\bibfield  {journal} {\bibinfo
			{journal} {Opt. Express}\ }\textbf {\bibinfo {volume} {12}},\ \bibinfo
		{pages} {4080} (\bibinfo {year} {2004})}\BibitemShut {NoStop}%
	\bibitem [{\citenamefont {Wen}\ \emph {et~al.}(2023)\citenamefont {Wen},
		\citenamefont {Dong}, \citenamefont {Deng}, \citenamefont {Pang},
		\citenamefont {Kaminski}, \citenamefont {Xu}, \citenamefont {Yan},
		\citenamefont {Wang}, \citenamefont {Liu}, \citenamefont {Tang},
		\citenamefont {Chen}, \citenamefont {Liu},\ and\ \citenamefont
		{Yang}}]{Wen_2023}%
	\BibitemOpen
	\bibfield  {author} {\bibinfo {author} {\bibfnamefont {Z.}~\bibnamefont
			{Wen}}, \bibinfo {author} {\bibfnamefont {Z.}~\bibnamefont {Dong}}, \bibinfo
		{author} {\bibfnamefont {Q.}~\bibnamefont {Deng}}, \bibinfo {author}
		{\bibfnamefont {C.}~\bibnamefont {Pang}}, \bibinfo {author} {\bibfnamefont
			{C.~F.}\ \bibnamefont {Kaminski}}, \bibinfo {author} {\bibfnamefont
			{X.}~\bibnamefont {Xu}}, \bibinfo {author} {\bibfnamefont {H.}~\bibnamefont
			{Yan}}, \bibinfo {author} {\bibfnamefont {L.}~\bibnamefont {Wang}}, \bibinfo
		{author} {\bibfnamefont {S.}~\bibnamefont {Liu}}, \bibinfo {author}
		{\bibfnamefont {J.}~\bibnamefont {Tang}}, \bibinfo {author} {\bibfnamefont
			{W.}~\bibnamefont {Chen}}, \bibinfo {author} {\bibfnamefont {X.}~\bibnamefont
			{Liu}},\ and\ \bibinfo {author} {\bibfnamefont {Q.}~\bibnamefont {Yang}},\
	}\href {https://doi.org/10.1038/s41566-023-01240-x} {\bibfield  {journal}
		{\bibinfo  {journal} {Nat. Photonics}\ }\textbf {\bibinfo {volume} {17}},\
		\bibinfo {pages} {679} (\bibinfo {year} {2023})}\BibitemShut {NoStop}%
	\bibitem [{\citenamefont {Lalanne}\ \emph {et~al.}(2018)\citenamefont
		{Lalanne}, \citenamefont {Yan}, \citenamefont {Vynck}, \citenamefont
		{Sauvan},\ and\ \citenamefont {Hugonin}}]{Lalanne_QNMReview_2018}%
	\BibitemOpen
	\bibfield  {author} {\bibinfo {author} {\bibfnamefont {P.}~\bibnamefont
			{Lalanne}}, \bibinfo {author} {\bibfnamefont {W.}~\bibnamefont {Yan}},
		\bibinfo {author} {\bibfnamefont {K.}~\bibnamefont {Vynck}}, \bibinfo
		{author} {\bibfnamefont {C.}~\bibnamefont {Sauvan}},\ and\ \bibinfo {author}
		{\bibfnamefont {J.-P.}\ \bibnamefont {Hugonin}},\ }\href
	{https://doi.org/10.1002/lpor.201700113} {\bibfield  {journal} {\bibinfo
			{journal} {Laser Photonics Rev.}\ }\textbf {\bibinfo {volume} {12}},\
		\bibinfo {pages} {1700113} (\bibinfo {year} {2018})}\BibitemShut {NoStop}%
	\bibitem [{\citenamefont {Lalanne}\ \emph {et~al.}(2019)\citenamefont
		{Lalanne}, \citenamefont {Yan}, \citenamefont {Gras}, \citenamefont {Sauvan},
		\citenamefont {Hugonin}, \citenamefont {Besbes}, \citenamefont {Dem\'{e}sy},
		\citenamefont {Truong}, \citenamefont {Gralak}, \citenamefont {Zolla},
		\citenamefont {Nicolet}, \citenamefont {Binkowski}, \citenamefont
		{Zschiedrich}, \citenamefont {Burger}, \citenamefont {Zimmerling},
		\citenamefont {Remis}, \citenamefont {Urbach}, \citenamefont {Liu},\ and\
		\citenamefont {Weiss}}]{Lalanne_QNM_Benchmark_2018}%
	\BibitemOpen
	\bibfield  {author} {\bibinfo {author} {\bibfnamefont {P.}~\bibnamefont
			{Lalanne}}, \bibinfo {author} {\bibfnamefont {W.}~\bibnamefont {Yan}},
		\bibinfo {author} {\bibfnamefont {A.}~\bibnamefont {Gras}}, \bibinfo {author}
		{\bibfnamefont {C.}~\bibnamefont {Sauvan}}, \bibinfo {author} {\bibfnamefont
			{J.-P.}\ \bibnamefont {Hugonin}}, \bibinfo {author} {\bibfnamefont
			{M.}~\bibnamefont {Besbes}}, \bibinfo {author} {\bibfnamefont
			{G.}~\bibnamefont {Dem\'{e}sy}}, \bibinfo {author} {\bibfnamefont {M.~D.}\
			\bibnamefont {Truong}}, \bibinfo {author} {\bibfnamefont {B.}~\bibnamefont
			{Gralak}}, \bibinfo {author} {\bibfnamefont {F.}~\bibnamefont {Zolla}},
		\bibinfo {author} {\bibfnamefont {A.}~\bibnamefont {Nicolet}}, \bibinfo
		{author} {\bibfnamefont {F.}~\bibnamefont {Binkowski}}, \bibinfo {author}
		{\bibfnamefont {L.}~\bibnamefont {Zschiedrich}}, \bibinfo {author}
		{\bibfnamefont {S.}~\bibnamefont {Burger}}, \bibinfo {author} {\bibfnamefont
			{J.}~\bibnamefont {Zimmerling}}, \bibinfo {author} {\bibfnamefont
			{R.}~\bibnamefont {Remis}}, \bibinfo {author} {\bibfnamefont
			{P.}~\bibnamefont {Urbach}}, \bibinfo {author} {\bibfnamefont {H.~T.}\
			\bibnamefont {Liu}},\ and\ \bibinfo {author} {\bibfnamefont {T.}~\bibnamefont
			{Weiss}},\ }\href {https://doi.org/10.1364/JOSAA.36.000686} {\bibfield
		{journal} {\bibinfo  {journal} {J. Opt. Soc. Am. A}\ }\textbf {\bibinfo
			{volume} {36}},\ \bibinfo {pages} {686} (\bibinfo {year} {2019})}\BibitemShut
	{NoStop}%
	\bibitem [{\citenamefont {Demésy}\ \emph {et~al.}(2020)\citenamefont
		{Demésy}, \citenamefont {Nicolet}, \citenamefont {Gralak}, \citenamefont
		{Geuzaine}, \citenamefont {Campos},\ and\ \citenamefont
		{Roman}}]{Demesy_ComputPhysComm_2020}%
	\BibitemOpen
	\bibfield  {author} {\bibinfo {author} {\bibfnamefont {G.}~\bibnamefont
			{Demésy}}, \bibinfo {author} {\bibfnamefont {A.}~\bibnamefont {Nicolet}},
		\bibinfo {author} {\bibfnamefont {B.}~\bibnamefont {Gralak}}, \bibinfo
		{author} {\bibfnamefont {C.}~\bibnamefont {Geuzaine}}, \bibinfo {author}
		{\bibfnamefont {C.}~\bibnamefont {Campos}},\ and\ \bibinfo {author}
		{\bibfnamefont {J.~E.}\ \bibnamefont {Roman}},\ }\href
	{https://doi.org/10.1016/j.cpc.2020.107509} {\bibfield  {journal} {\bibinfo
			{journal} {Comput. Phys. Commun.}\ }\textbf {\bibinfo {volume} {257}},\
		\bibinfo {pages} {107509} (\bibinfo {year} {2020})}\BibitemShut {NoStop}%
	\bibitem [{\citenamefont {Birks}\ \emph {et~al.}(1997)\citenamefont {Birks},
		\citenamefont {Knight},\ and\ \citenamefont {{P. St. J.
				Russell}}}]{Birks_1997}%
	\BibitemOpen
	\bibfield  {author} {\bibinfo {author} {\bibfnamefont {T.~A.}\ \bibnamefont
			{Birks}}, \bibinfo {author} {\bibfnamefont {J.~C.}\ \bibnamefont {Knight}},\
		and\ \bibinfo {author} {\bibnamefont {{P. St. J. Russell}}},\ }\href
	{https://doi.org/10.1364/OL.22.000961} {\bibfield  {journal} {\bibinfo
			{journal} {Opt. Lett.}\ }\textbf {\bibinfo {volume} {22}},\ \bibinfo {pages}
		{961} (\bibinfo {year} {1997})}\BibitemShut {NoStop}%
	\bibitem [{\citenamefont {White}\ \emph {et~al.}(2002)\citenamefont {White},
		\citenamefont {McPhedran}, \citenamefont {de~Sterke}, \citenamefont
		{Litchinitser},\ and\ \citenamefont {Eggleton}}]{White_OptLett_2002}%
	\BibitemOpen
	\bibfield  {author} {\bibinfo {author} {\bibfnamefont {T.~P.}\ \bibnamefont
			{White}}, \bibinfo {author} {\bibfnamefont {R.~C.}\ \bibnamefont
			{McPhedran}}, \bibinfo {author} {\bibfnamefont {C.~M.}\ \bibnamefont
			{de~Sterke}}, \bibinfo {author} {\bibfnamefont {N.~M.}\ \bibnamefont
			{Litchinitser}},\ and\ \bibinfo {author} {\bibfnamefont {B.~J.}\ \bibnamefont
			{Eggleton}},\ }\href {https://doi.org/10.1364/OL.27.001977} {\bibfield
		{journal} {\bibinfo  {journal} {Opt. Lett.}\ }\textbf {\bibinfo {volume}
			{27}},\ \bibinfo {pages} {1977} (\bibinfo {year} {2002})}\BibitemShut
	{NoStop}%
	\bibitem [{\citenamefont {Eggleton}\ \emph {et~al.}(2000)\citenamefont
		{Eggleton}, \citenamefont {Westbrook}, \citenamefont {White}, \citenamefont
		{Kerbage}, \citenamefont {Windeler},\ and\ \citenamefont
		{Burdge}}]{Eggleton_2000}%
	\BibitemOpen
	\bibfield  {author} {\bibinfo {author} {\bibfnamefont {B.~J.}\ \bibnamefont
			{Eggleton}}, \bibinfo {author} {\bibfnamefont {P.~S.}\ \bibnamefont
			{Westbrook}}, \bibinfo {author} {\bibfnamefont {C.~A.}\ \bibnamefont
			{White}}, \bibinfo {author} {\bibfnamefont {C.}~\bibnamefont {Kerbage}},
		\bibinfo {author} {\bibfnamefont {R.~S.}\ \bibnamefont {Windeler}},\ and\
		\bibinfo {author} {\bibfnamefont {G.~L.}\ \bibnamefont {Burdge}},\
	}\href@noop {} {\bibfield  {journal} {\bibinfo  {journal} {J. Lightwave
				Technol.}\ }\textbf {\bibinfo {volume} {18}},\ \bibinfo {pages} {1084}
		(\bibinfo {year} {2000})}\BibitemShut {NoStop}%
	\bibitem [{\citenamefont {Uebel}\ \emph {et~al.}(2016)\citenamefont {Uebel},
		\citenamefont {G\"{u}nendi}, \citenamefont {Frosz}, \citenamefont {Ahmed},
		\citenamefont {Edavalath}, \citenamefont {M\'{e}nard},\ and\ \citenamefont
		{{P. St. J. Russell}}}]{Uebel_OL_2016}%
	\BibitemOpen
	\bibfield  {author} {\bibinfo {author} {\bibfnamefont {P.}~\bibnamefont
			{Uebel}}, \bibinfo {author} {\bibfnamefont {M.~C.}\ \bibnamefont
			{G\"{u}nendi}}, \bibinfo {author} {\bibfnamefont {M.~H.}\ \bibnamefont
			{Frosz}}, \bibinfo {author} {\bibfnamefont {G.}~\bibnamefont {Ahmed}},
		\bibinfo {author} {\bibfnamefont {N.~N.}\ \bibnamefont {Edavalath}}, \bibinfo
		{author} {\bibfnamefont {J.-M.}\ \bibnamefont {M\'{e}nard}},\ and\ \bibinfo
		{author} {\bibnamefont {{P. St. J. Russell}}},\ }\href
	{https://doi.org/10.1364/OL.41.001961} {\bibfield  {journal} {\bibinfo
			{journal} {Opt. Lett.}\ }\textbf {\bibinfo {volume} {41}},\ \bibinfo {pages}
		{1961} (\bibinfo {year} {2016})}\BibitemShut {NoStop}%
	\bibitem [{\citenamefont {Roberts}\ \emph {et~al.}(2005)\citenamefont
		{Roberts}, \citenamefont {Couny}, \citenamefont {Sabert}, \citenamefont
		{Mangan}, \citenamefont {Williams}, \citenamefont {Farr}, \citenamefont
		{Mason}, \citenamefont {Tomlinson}, \citenamefont {Birks}, \citenamefont
		{Knight},\ and\ \citenamefont {{P. St. J. Russell}}}]{Roberts_2005}%
	\BibitemOpen
	\bibfield  {author} {\bibinfo {author} {\bibfnamefont {P.~J.}\ \bibnamefont
			{Roberts}}, \bibinfo {author} {\bibfnamefont {F.}~\bibnamefont {Couny}},
		\bibinfo {author} {\bibfnamefont {H.}~\bibnamefont {Sabert}}, \bibinfo
		{author} {\bibfnamefont {B.~J.}\ \bibnamefont {Mangan}}, \bibinfo {author}
		{\bibfnamefont {D.~P.}\ \bibnamefont {Williams}}, \bibinfo {author}
		{\bibfnamefont {L.}~\bibnamefont {Farr}}, \bibinfo {author} {\bibfnamefont
			{M.~W.}\ \bibnamefont {Mason}}, \bibinfo {author} {\bibfnamefont
			{A.}~\bibnamefont {Tomlinson}}, \bibinfo {author} {\bibfnamefont {T.~A.}\
			\bibnamefont {Birks}}, \bibinfo {author} {\bibfnamefont {J.~C.}\ \bibnamefont
			{Knight}},\ and\ \bibinfo {author} {\bibnamefont {{P. St. J. Russell}}},\
	}\href {https://doi.org/10.1364/OPEX.13.000236} {\bibfield  {journal}
		{\bibinfo  {journal} {Opt. Express}\ }\textbf {\bibinfo {volume} {13}},\
		\bibinfo {pages} {236} (\bibinfo {year} {2005})}\BibitemShut {NoStop}%
	\bibitem [{\citenamefont {Pearce}\ \emph {et~al.}(2007)\citenamefont {Pearce},
		\citenamefont {Wiederhecker}, \citenamefont {Poulton}, \citenamefont
		{Burger},\ and\ \citenamefont {{P. St. J. Russell}}}]{Pearce2007oe}%
	\BibitemOpen
	\bibfield  {author} {\bibinfo {author} {\bibfnamefont {G.~J.}\ \bibnamefont
			{Pearce}}, \bibinfo {author} {\bibfnamefont {G.~S.}\ \bibnamefont
			{Wiederhecker}}, \bibinfo {author} {\bibfnamefont {C.~G.}\ \bibnamefont
			{Poulton}}, \bibinfo {author} {\bibfnamefont {S.}~\bibnamefont {Burger}},\
		and\ \bibinfo {author} {\bibnamefont {{P. St. J. Russell}}},\ }\href
	{https://doi.org/10.1364/OE.15.012680} {\bibfield  {journal} {\bibinfo
			{journal} {Opt. Express}\ }\textbf {\bibinfo {volume} {15}},\ \bibinfo
		{pages} {12680} (\bibinfo {year} {2007})}\BibitemShut {NoStop}%
	\bibitem [{\citenamefont {Nakatsukasa}\ \emph {et~al.}(2018)\citenamefont
		{Nakatsukasa}, \citenamefont {S\`{e}te},\ and\ \citenamefont
		{Trefethen}}]{Nakatsukasa_2018_AAA}%
	\BibitemOpen
	\bibfield  {author} {\bibinfo {author} {\bibfnamefont {Y.}~\bibnamefont
			{Nakatsukasa}}, \bibinfo {author} {\bibfnamefont {O.}~\bibnamefont
			{S\`{e}te}},\ and\ \bibinfo {author} {\bibfnamefont {L.~N.}\ \bibnamefont
			{Trefethen}},\ }\href {https://doi.org/10.1137/16M1106122} {\bibfield
		{journal} {\bibinfo  {journal} {SIAM J. Sci. Comput.}\ }\textbf {\bibinfo
			{volume} {40}},\ \bibinfo {pages} {A1494} (\bibinfo {year}
		{2018})}\BibitemShut {NoStop}%
	\bibitem [{\citenamefont {Elsworth}\ and\ \citenamefont
		{Güttel}(2019)}]{Elsworth_2019}%
	\BibitemOpen
	\bibfield  {author} {\bibinfo {author} {\bibfnamefont {S.}~\bibnamefont
			{Elsworth}}\ and\ \bibinfo {author} {\bibfnamefont {S.}~\bibnamefont
			{Güttel}},\ }\href {https://doi.org/10.1016/j.laa.2018.10.003} {\bibfield
		{journal} {\bibinfo  {journal} {Linear Algebra Its Appl.}\ }\textbf {\bibinfo
			{volume} {576}},\ \bibinfo {pages} {246} (\bibinfo {year}
		{2019})}\BibitemShut {NoStop}%
	\bibitem [{\citenamefont {Lietaert}\ \emph {et~al.}(2021)\citenamefont
		{Lietaert}, \citenamefont {Meerbergen}, \citenamefont {Pérez},\ and\
		\citenamefont {Vandereycken}}]{Lietaert_2021}%
	\BibitemOpen
	\bibfield  {author} {\bibinfo {author} {\bibfnamefont {P.}~\bibnamefont
			{Lietaert}}, \bibinfo {author} {\bibfnamefont {K.}~\bibnamefont
			{Meerbergen}}, \bibinfo {author} {\bibfnamefont {J.}~\bibnamefont {Pérez}},\
		and\ \bibinfo {author} {\bibfnamefont {B.}~\bibnamefont {Vandereycken}},\
	}\href {https://doi.org/10.1093/imanum/draa098} {\bibfield  {journal}
		{\bibinfo  {journal} {IMA J. Numer. Anal.}\ }\textbf {\bibinfo {volume}
			{42}},\ \bibinfo {pages} {1087} (\bibinfo {year} {2021})}\BibitemShut
	{NoStop}%
	\bibitem [{\citenamefont {G\"{u}ttel}\ \emph {et~al.}(2022)\citenamefont
		{G\"{u}ttel}, \citenamefont {Negri~Porzio},\ and\ \citenamefont
		{Tisseur}}]{Guettel_weightedAAA_2022}%
	\BibitemOpen
	\bibfield  {author} {\bibinfo {author} {\bibfnamefont {S.}~\bibnamefont
			{G\"{u}ttel}}, \bibinfo {author} {\bibfnamefont {G.~M.}\ \bibnamefont
			{Negri~Porzio}},\ and\ \bibinfo {author} {\bibfnamefont {F.}~\bibnamefont
			{Tisseur}},\ }\href {https://doi.org/10.1137/20M1380533} {\bibfield
		{journal} {\bibinfo  {journal} {SIAM J. Sci. Comput.}\ }\textbf {\bibinfo
			{volume} {44}},\ \bibinfo {pages} {A2439} (\bibinfo {year}
		{2022})}\BibitemShut {NoStop}%
	\bibitem [{\citenamefont {Bruno}\ \emph {et~al.}(2024)\citenamefont {Bruno},
		\citenamefont {Santana},\ and\ \citenamefont {Trefethen}}]{Bruno_AAA_2024}%
	\BibitemOpen
	\bibfield  {author} {\bibinfo {author} {\bibfnamefont {O.~P.}\ \bibnamefont
			{Bruno}}, \bibinfo {author} {\bibfnamefont {M.~A.}\ \bibnamefont {Santana}},\
		and\ \bibinfo {author} {\bibfnamefont {L.~N.}\ \bibnamefont {Trefethen}},\
	}\bibfield  {journal} {\bibinfo  {journal} {preprint arXiv:2405.19582v2}\
	}\href {https://doi.org/10.48550/arXiv.2405.19582}
	{10.48550/arXiv.2405.19582} (\bibinfo {year} {2024})\BibitemShut {NoStop}%
	\bibitem [{\citenamefont {Betz}\ \emph {et~al.}(2024)\citenamefont {Betz},
		\citenamefont {Hammerschmidt}, \citenamefont {Zschiedrich}, \citenamefont
		{Burger},\ and\ \citenamefont {Binkowski}}]{Betz_LPOR_2024}%
	\BibitemOpen
	\bibfield  {author} {\bibinfo {author} {\bibfnamefont {F.}~\bibnamefont
			{Betz}}, \bibinfo {author} {\bibfnamefont {M.}~\bibnamefont {Hammerschmidt}},
		\bibinfo {author} {\bibfnamefont {L.}~\bibnamefont {Zschiedrich}}, \bibinfo
		{author} {\bibfnamefont {S.}~\bibnamefont {Burger}},\ and\ \bibinfo {author}
		{\bibfnamefont {F.}~\bibnamefont {Binkowski}},\ }\href
	{https://doi.org/10.1002/lpor.202400584} {\bibfield  {journal} {\bibinfo
			{journal} {Laser Photonics Rev.}\ }\textbf {\bibinfo {volume} {18}},\
		\bibinfo {pages} {2400584} (\bibinfo {year} {2024})}\BibitemShut {NoStop}%
	\bibitem [{\citenamefont {Pradovera}(2023)}]{Pradovera_PAMM_2023}%
	\BibitemOpen
	\bibfield  {author} {\bibinfo {author} {\bibfnamefont {D.}~\bibnamefont
			{Pradovera}},\ }\href {https://doi.org/10.1002/pamm.202200032} {\bibfield
		{journal} {\bibinfo  {journal} {PAMM}\ }\textbf {\bibinfo {volume} {22}},\
		\bibinfo {pages} {e202200032} (\bibinfo {year} {2023})}\BibitemShut {NoStop}%
	\bibitem [{\citenamefont {Binkowski}\ \emph {et~al.}(2024)\citenamefont
		{Binkowski}, \citenamefont {Betz}, \citenamefont {Hammerschmidt},
		\citenamefont {Zschiedrich},\ and\ \citenamefont
		{Burger}}]{Binkowski_SourceCode_AAA_HCPCF}%
	\BibitemOpen
	\bibfield  {author} {\bibinfo {author} {\bibfnamefont {F.}~\bibnamefont
			{Binkowski}}, \bibinfo {author} {\bibfnamefont {F.}~\bibnamefont {Betz}},
		\bibinfo {author} {\bibfnamefont {M.}~\bibnamefont {Hammerschmidt}}, \bibinfo
		{author} {\bibfnamefont {L.}~\bibnamefont {Zschiedrich}},\ and\ \bibinfo
		{author} {\bibfnamefont {S.}~\bibnamefont {Burger}},\ }\href
	{https://doi.org/10.5281/zenodo.14202409} {\bibinfo {title} {{Source code and
				simulation results: Resonance modes in microstructured photonic waveguides -
				Efficient and accurate computation based on AAA rational approximation}}},\
	\bibinfo {howpublished} {Zenodo} (\bibinfo {year} {2024}),\ \bibinfo {note}
	{doi: 10.5281/zenodo.14202409}\BibitemShut {NoStop}%
	\bibitem [{\citenamefont {Saad}(2011)}]{Saad_Book_NumMeth_Eig_2011}%
	\BibitemOpen
	\bibfield  {author} {\bibinfo {author} {\bibfnamefont {Y.}~\bibnamefont
			{Saad}},\ }\href@noop {} {\emph {\bibinfo {title} {{Numerical Methods for
					Large Eigenvalue Problems, 2nd ed.}}}}\ (\bibinfo  {publisher} {SIAM},\
	\bibinfo {address} {Philadelphia},\ \bibinfo {year} {2011})\BibitemShut
	{NoStop}%
	\bibitem [{\citenamefont {Binkowski}\ \emph {et~al.}(2022)\citenamefont
		{Binkowski}, \citenamefont {Betz}, \citenamefont {Hammerschmidt},
		\citenamefont {Schneider}, \citenamefont {Zschiedrich},\ and\ \citenamefont
		{Burger}}]{Binkowski_CommunPhys_2022}%
	\BibitemOpen
	\bibfield  {author} {\bibinfo {author} {\bibfnamefont {F.}~\bibnamefont
			{Binkowski}}, \bibinfo {author} {\bibfnamefont {F.}~\bibnamefont {Betz}},
		\bibinfo {author} {\bibfnamefont {M.}~\bibnamefont {Hammerschmidt}}, \bibinfo
		{author} {\bibfnamefont {P.-I.}\ \bibnamefont {Schneider}}, \bibinfo {author}
		{\bibfnamefont {L.}~\bibnamefont {Zschiedrich}},\ and\ \bibinfo {author}
		{\bibfnamefont {S.}~\bibnamefont {Burger}},\ }\href
	{https://doi.org/10.1038/s42005-022-00977-1} {\bibfield  {journal} {\bibinfo
			{journal} {Commun. Phys.}\ }\textbf {\bibinfo {volume} {5}},\ \bibinfo
		{pages} {202} (\bibinfo {year} {2022})}\BibitemShut {NoStop}%
\end{thebibliography}
\end{document}